# Explainable histomorphology-based survival prediction of glioblastoma, IDH-wildtype

Jan-Philipp Redlich[1,*], Friedrich Feuerhake[2,11], Stefan Nikolin[3], Nadine Sarah Schaadt[2], Sarah Teuber-Hanselmann[4], Joachim Weis[3], Sabine Luttmann[5], Andrea Eberle[5], Christoph Buck[6], Timm Intemann[6], Pascal Birnstill[7], Klaus Kraywinkel[8], Jonas Ort[9,12], Peter Boor[10], André Homeyer[1]

[1] Fraunhofer Institute for Digital Medicine MEVIS, Max-von-Laue-Straße 2, 28359, Bremen, Germany
[2] Hannover Medical School, Carl-Neuberg-Straße 1, 30625 Hannover, Germany
[3] Institute of Neuropathology, RWTH Aachen University Hospital, Pauwelsstraße 30, 52074, Aachen, Germany
[4] Department of Neuropathology, Center for Pathology, Klinikum Bremen-Mitte, Sankt-Jürgen-Straße 1, 28205 Bremen, Germany
[5] Bremen Cancer Registry, Leibniz Institute for Prevention Research and Epidemiology - BIPS, Achterstraße 30, 28359 Bremen, Germany
[6] Leibniz Institute for Prevention Research and Epidemiology - BIPS, Achterstraße 30, 28359 Bremen, Germany
[7] Fraunhofer Institute of Optronics, System Technologies and Image Exploitation IOSB, Fraunhoferstraße 1, 76131 Karlsruhe, Germany
[8] Robert Koch Institute, Nordufer 20, 13353 Berlin, Germany
[9] Department of Neurosurgery, RWTH Aachen University Hospital, Pauwelsstraße 30, 52074, Aachen, Germany
[10] Institute of Pathology, RWTH Aachen University Hospital, Pauwelsstraße 30, 52074, Aachen, Germany
[11] Institute of Neuropathology, Medical Center - University of Freiburg, Breisacher Str. 64, 79106 Freiburg, Germany
[12] Center for Integrated Oncology Aachen Bonn Cologne Duesseldorf (CIO ABCD), Germany

*Corresponding author. E-mail address: jan.ph.redlich@gmail.com

## Abstract

Glioblastoma, IDH-wildtype (GBM-IDHwt) is the most common malignant brain tumor. While histomorphology is a crucial component of GBM-IDHwt diagnosis, it is not further considered for prognosis. Here, we present an explainable artificial intelligence (AI) framework to identify and interpret histomorphological features associated with patient survival. The framework combines an explainable multiple instance learning (MIL) architecture that directly identifies prognostically relevant image tiles with a sparse autoencoder (SAE) that maps these tiles to interpretable visual patterns. The MIL model was trained and evaluated on a new real-world dataset of 720 GBM-IDHwt cases from three hospitals and four cancer registries across Germany. The SAE was trained on 1,878 whole-slide images from five independent public glioblastoma collections. Despite the many factors influencing survival time, our method showed some ability to discriminate between patients living less than 180 days or more than 360 days solely based on histomorphology (AUC: 0.67; 95% CI: 0.63–0.72). Cox proportional hazards regression confirmed a significant survival difference between predicted groups after adjustment for established prognostic factors (hazard ratio: 1.47; 95% CI: 1.26–1.72). Three neuropathologists categorized the identified visual patterns into seven distinct histomorphological groups, revealing both established prognostic features and unexpected associations, the latter being potentially attributable to surgery-related confounders. The presented explainable AI framework facilitates prognostic biomarker discovery in GBM-IDHwt and beyond, highlighting promising histomorphological features for further analysis and exposing potential confounders that would be hidden in black-box models.

# 1. Introduction

Glioblastoma is the most common primary malignant brain tumor in adults[1]. Since the 2021 World Health Organization (WHO) classification, the term "glioblastoma" has been limited to grade 4 gliomas with isocitrate dehydrogenase (IDH)-wildtype status (GBM-IDHwt). Astrocytic IDH-mutant malignant (grade 4) gliomas, which were previously also considered glioblastomas, are now recognized as a separate entity with a more favorable prognosis[2]. GBM-IDHwt has a poor prognosis, with a median overall survival of roughly 12 months despite treatment, and only a small proportion of patients survive beyond five years[3,4]. Its incidence is about 2 per 100,000 person-years and it is more common in men[1,5].

GBM-IDHwt is diagnosed using tissue obtained through surgical resection or biopsy. Diagnosis integrates histomorphological features from hematoxylin and eosin (H&E) stained tissue sections and molecular features from immunohistochemistry or gene sequencing. Histologically, GBM-IDHwt are characterized by pleomorphic, often astrocyte-like, diffusely infiltrating tumor cells, often with microvascular proliferation or necrosis. On the molecular level, beyond IDH-wildtype status, hallmark alterations include TERT promoter mutations, EGFR amplification, or a combined gain of chromosome 7 and loss of chromosome 10[2]. Standard treatment of glioblastoma follows the Stupp protocol, which consists of maximal safe tumor resection followed by radiotherapy plus temozolomide chemotherapy[6].

Accurate prognostication can be essential for the clinical management of GBM-IDHwt. For example, it can help to plan the extent of secondary neurosurgical interventions, personalize radiation therapy and chemotherapy protocols, and support patient care, counseling, and inclusion in clinical trials. However, the short survival time and substantial histomorphological and molecular heterogeneity of GBM-IDHwt make prognostication difficult. The most commonly considered prognostic factors are age at diagnosis, Karnofsky performance status, MGMT promoter methylation, tumor location, and extent of resection[7–10]. Beyond diagnosis, histomorphological features are currently not considered for further prognosis.

## 1.1 AI-based prognosis

Artificial intelligence (AI) methods have shown promise to automatically extract prognostic information from histological whole-slide images (WSI) of glioblastoma tissue[11]. The proposed methods either predict risk scores or survival times, or classify patients into discrete survival groups[11]. Recent approaches typically divide WSIs into tiles and involve two main processing components: a foundation model (FM), which encodes the tiles as numerical vector representations, so called embeddings, and a multiple instance learning (MIL) method, which aggregates the embeddings and infers a single prediction for the entire WSI.

FMs for pathology are a major interest of current research. These large-scale deep neural networks were trained using extensive datasets spanning several hundred thousand WSIs from diverse tissue types and laboratories, enabling them to capture rich, general-purpose abstractions of the tile content[12]. They have demonstrated unprecedented performance in a wide variety of downstream tasks, surpassing smaller deep neural networks employed by previous studies and conventionally trained using natural images[12].

MIL architectures have the advantage that they can be trained using only WSI-level annotations, which are more abundant and easier to create, allowing them to benefit from larger amounts of data[13]. Various neural network architectures have been proposed for MIL, differing in how they aggregate and weight tile evidence.

## 1.2 Explainability

For clinical translation, it is crucial to determine which histomorphological features AI models have learned to associate with survival. This includes verifying that model predictions are based on meaningful features rather than on spurious correlations. Examples of survival-associated features of many cancer types include nuclear pleomorphism, hypercellularity, infiltrative growth, microvascular proliferation, or necrosis. Since AI models are data-driven and hypothesis-independent, they may uncover previously unknown survival-related morphologies, potentially refining prognostication in GBM-IDHwt.

Modern AI models are complex black boxes whose internal workings are difficult to interpret. Explainability methods can highlight the most influential factors behind a model's prediction in a human-understandable way[14,15]. Most MIL approaches employ attention mechanisms that enable the model to focus on the most informative image regions. Each tile is assigned an attention score quantifying its influence on the WSI-level prediction. Consequently, tiles with higher attention scores can be expected to capture the histomorphological features most strongly associated with survival. Visualizing the attention scores as colorized heatmaps overlaying the WSI allows pathologists to qualitatively assess the models' learned associations. To enable quantitative analysis of these associations, the tiles must first be mapped to discrete histomorphological features. FM embeddings offer a promising basis for this mapping, as they encode rich histomorphological information. However, further processing is required to uncover human-understandable histomorphological features from FM embeddings.

Cluster-based approaches are often used for feature mapping. They are based on the assumption that the distance between FM embeddings quantifies the visual similarity of the underlying tiles, and employ clustering algorithms, such as HDBSCAN[16], to group nearby embeddings representing similar visual patterns. The resulting clusters are then mapped to histomorphological features through human visual inspection[17]. These approaches have two drawbacks. First, meaningful clustering necessitates projecting high-dimensional embeddings into a low-dimensional space, e.g., using dimensionality reduction algorithms, such as UMAP[18]. In this low-dimensional space, proximity does not necessarily guarantee histomorphological similarity. Second, the derived histomorphological features depend strongly on the chosen dimensionality reduction and clustering algorithms, as well as their parameters. Adjusting these choices to make clusters reflect meaningful morphologies is difficult and prone to bias.

Sparse autoencoders (SAEs) have recently demonstrated an unparalleled ability to map FM embeddings into human-interpretable features across diverse domains[19–21]. In pathology, initial studies suggest that SAEs have the potential to support the identification of meaningful histomorphological features of different tissue types[22,23]. In this context, SAEs are trained using millions of tile embeddings to learn a set of visual patterns which allows each tile to be encoded as a sparse linear combination of these patterns. The number of the learned patterns greatly exceeds the dimensionality of the FM embedding space yet remains much smaller than the total number of tiles. Combined with the sparsity constraint, this causes SAEs to learn visual patterns that are interpretable for humans[20]. The mapping from learned visual patterns to meaningful histomorphological features still has to be performed by human visual inspection.

## 1.3 Contributions

Only a few previous studies have specifically considered the disease entity GBM-IDHwt[17,24–27]. Others were based on previously defined diagnostic criteria or combined multiple disease entities from the broader category of adult-type diffuse glioma. A review is given by Redlich et al.[11]. The GBM-IDHwt-specific studies were based on small datasets, primarily from two publicly available data collections originally compiled for different purposes[28–30]. This limits the potential of these datasets to represent the real-world case spectrum of GBM-IDHwt and their generalizability to practical applications[11,31].

Moreover, only a few of the related studies have investigated which histomorphological features were learned to be associated with survival[17,25]. This severely impairs the clinical translatability of the results. When such investigations were conducted, they relied on manual searching for notable differences between longer and shorter survivors, either in attention heatmaps of selected cases or in visualizations of clustered, low-dimensional tile embeddings. This qualitative approach is potentially biased by observer expectations and complicates quantitative analysis of associations with survival. Overall, the reported associations remain largely inconclusive.

To address this, we propose an explainable AI framework to identify and interpret histomorphological features associated with patient survival (Figure 1). By combining an explainable MIL architecture and a SAE, the framework enables unbiased quantitative analysis of the learned prognostic associations. We evaluated this framework on a large real-world dataset of GBM-IDHwt specifically compiled for this study and present an interpretation of the results by three neuropathologists.

# 2. Materials and Methods

Our study comprised six steps that are explained in more detail in the following subsections (Figure 1). Steps two to five constitute our explainable AI-based method and were fully automated. Step 1: The training and evaluation dataset was collected by linking pathology and cancer registry data and prepared for analysis. Step 2: The explainable MIL architecture was trained and its prognostic value was evaluated. The architecture quantified the association of each tile with different survival groups using decision scores. Step 3: Prognosis-relevant tiles were sampled based on the tile-based decision scores. Step 4: Using a SAE, the sampled tiles were mapped to human-interpretable visual patterns based on their FM embeddings. Prior to application, the SAE was trained on independent datasets to learn GBM-IDHwt-specific visual patterns. Step 5: Relevant visual patterns were selected based on their SAE activations among the sampled tiles. Step 6: Three board-certified neuropathologists with longstanding experience in diagnostic evaluation of brain tumors interpreted the selected visual patterns to identify histomorphological features. The code implementing the framework is released open-source at https://github.com/FraunhoferMEVIS/canconnect-study.

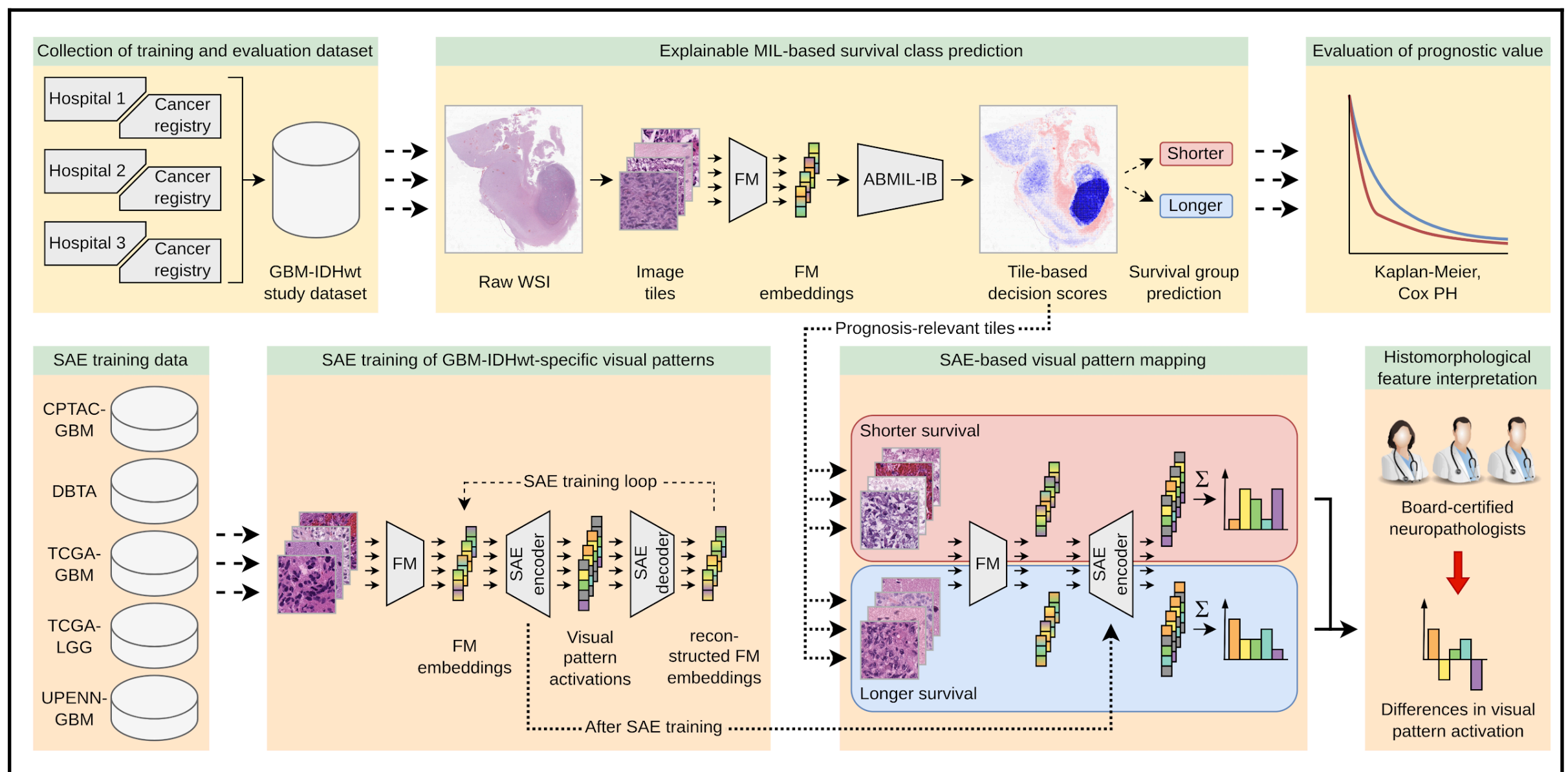

**Figure 1:** Illustration of the study design. The top row describes the explainable MIL-based survival group prediction and evaluation of prognostic value using the collected training and evaluation dataset. The bottom row describes the SAE training, the SAE-based visual pattern mapping and interpretation of prognosis-relevant histomorphological features by neuropathologists.

## 2.1 Data collection

A new glioblastoma dataset was compiled based on real-world clinical data. The dataset comprised consecutive patients diagnosed by the neuropathology service of three German hospitals, including two University Medical Centers, and one clinic of corresponding maximal level of care, affiliated with a University Medical Center. All three hospitals are diagnostic referral centers, providing diagnostic service for in-house brain tumor patients, as well as for patients who undergo subsequent treatment in different medical institutions. The dataset consisted of routinely collected pathology data from the neuropathology institutes of the respective hospitals. This includes WSIs of primary glioblastoma tumor samples and molecular biomarker information. Additional demographic, therapy and survival information about the patients was supplemented by linking the pathology data with cancer registry data. A privacy-preserving data linkage process was employed to compile and combine the pathology and cancer registry data (Supplementary Figure 1).

German hospitals are required by law to report significant events in the diagnosis and treatment of cancer patients to a local cancer registry[32]. There are more than fifteen cancer registries in Germany, each of which is responsible for different federal states or specializes in population-based or clinical data collection[33]. These registers curate the reported data and supplement death information from civil registry offices. Researchers can request anonymized data subsets from the cancer registries[34]. Table 1 lists the three hospitals that provided the pathology data, as well as the cancer registries that provided the respective cancer registry data.

| | **Hospital 1** | **Hospital 2** | **Hospital 3** |
|---|---|---|---|
| **Federal state** | Bremen | Lower Saxony | North Rhine-Westphalia |
| **Hospital** | Klinikum Bremen-Mitte | Medizinische Hochschule Hannover | Universitätsklinikum Aachen |
| **Cancer registries** | Bremer Krebsregister | Klinisches Krebsregister Niedersachsen, Epidemiologisches Krebsregister Niedersachsen | Landeskrebsregister Nordrhein-Westfalen |

**Table 1:** Overview of the three hospitals that provided pathology data, as well as the cancer registries that provided the respective cancer registry data.

Patients were included in the dataset if they were (1) diagnosed with primary GBM-IDHwt at one of the three hospitals, (2) at least 18 years old at the time of diagnosis, and (3) diagnosed between the years 2016 and 2023. Primary GBM-IDHwt cases were defined as those with an ICD-O-3 diagnosis code of 9440/3, 9441/3, or 9442/3 and IDH-wildtype status. Follow-up ended in June 2025. Patients were excluded in case of inconclusive survival information, missing information on the primary tumor when recurrent tumors were the first entry in the records, insufficient size or quality of tissue specimen for scanning, or inconsistencies between records of the neuropathology archives and cancer registries. Of the patients from Hospital 3, only those who lived in the same federal state could be considered, due to specific requirements of the respective cancer registry.

## 2.2 Data preparation

The dataset contained one whole-slide image per patient showing a tissue section from the primary tumor that was formalin-fixed, paraffin-embedded, and stained with hematoxylin and eosin (H&E). The sections were selected by trained neuropathologists. The tissue sections from Hospitals 1 and 2 were digitized at Hospital 2 using a Leica Aperio AT2 Scanner or a Leica Aperio GT450 Scanner. The tissue sections from Hospital 3 were digitized locally using a Leica Aperio AT2 Scanner. In addition to the image data, the pathology data included tabular information on the IDH mutation status and MGMT promoter methylation status of the patients that was extracted from the respective laboratory information systems.

All WSIs were processed into image tiles using the Slideflow library (version 3.0)[35]. Tissue was segmented from the slide background using Otsu thresholding; pen markings were removed using a Gaussian-blur-filtering-based method. Segmented tissue was partitioned into image tiles of size 224 x 224 pixels at a resolution of 0.5 µm/pixel, yielding 11.1 million patches in total (mean 15,430 per WSI). Each tile was embedded into a 2560-dimensional vector using the Virchow2 foundation model[36]. The first 1280 dimensions (the Vision Transformer class token) were used for downstream analysis. The Virchow2 foundation model was chosen over other computational pathology foundation models due to its extensive training data, which includes a relatively high proportion of brain tissue, and its comparably high robustness with respect to technical variability across hospitals[36,37].

The cancer registry data consisted of records of cancer cases in accordance with the German basic oncology dataset format (Onkologischer Basisdatensatz, or oBDS)[32]. For anonymization purposes, the cancer registries coarsened certain patient variables: Age was provided in five-year groups and dates (e.g., time of death) were provided as the number of days since diagnosis. Some remaining inconsistencies in the cancer registry data were identified and corrected in collaboration with the respective registries.

## 2.3 Explainable MIL architecture

With respect to the difficulty of prognosis of GBM-IDHwt, survival prediction was treated as a classification task based on two well-separated survival groups. The shorter-survival group included patients who died within 180 days after their initial diagnosis, excluding those who were censored before 180 days. The longer-survival group included patients who survived at least 360 days. To improve discriminability, patients with intermediate survival (>180 and <360 days) were excluded in the context of classification. The cut-off values were selected based on established clinical follow-up intervals.

We propose an explainable MIL method, named ABMIL-IB, to discriminate between patients with shorter and longer survival. The method builds on attention-based MIL (ABMIL), which is simple and has demonstrated strong performance across histopathology tasks[38–41]. ABMIL follows an embedding-based MIL approach, where tile embeddings are first aggregated into an intermediate WSI-level representation that is subsequently processed to yield the prediction. In contrast, our ABMIL-IB method adopts an instance-based approach. First, each tile is assigned a decision score indicating its association with either shorter or longer survival. Then, the scores of all tiles are normalized by the attention scores across the entire WSI and summed to obtain the WSI-level prediction[13]. This offers superior explainability compared to embedding-based approaches as it enables to unambiguously retrace the contribution of each tile to the final prediction (Figure 2).

The ABMIL-IB methods predicts the patient's probability of belonging to the longer-survival group as

$$P_{longer}(X) \ = \ sigmoid\left(\frac{1}{\sum_{j=1}^{n} exp(a(x_j))} \sum_{i=1}^{n} exp(a(x_i)) p(x_i)\right),$$

where $X = \{x_1, ..., x_n\}$ is the patient's WSI, represented by the FM embeddings $x_i$ of its tiles, and $a$ and $p$ are the attention and predictor neural networks, respectively. The final survival group prediction is inferred from $P_{longer}(X)$ using a decision threshold of 0.5. The tile-based decision scores are represented by the term $exp(a(x_i))p(x_i)$. The approach is similar to AdditiveMIL[42], but applies the softmax-weighted attention scores $\frac{exp(a(x_i))}{\sum_{j=1}^{n} exp(a(x_j))}$ directly to the predictions $p(x_i)$ instead of the embeddings $x_i$.

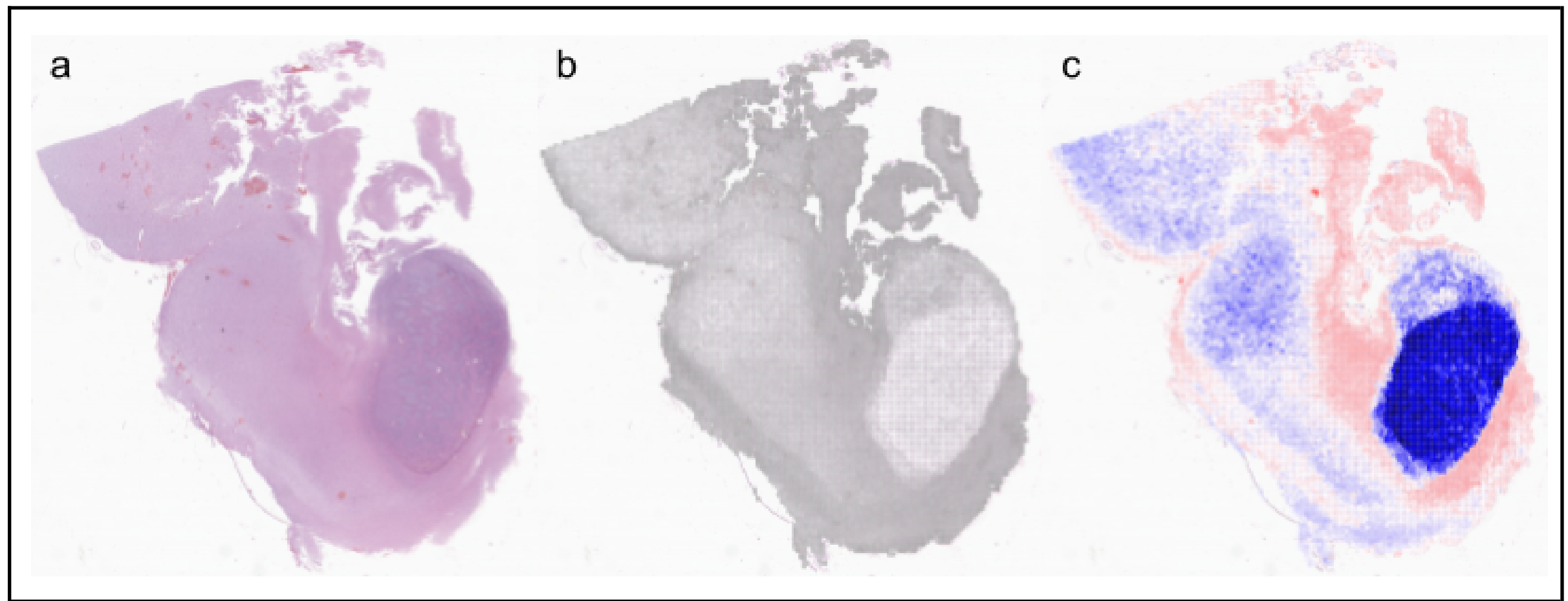

**Figure 2:** WSI of a patient from the longer-survival group of Hospital 2 and colorized heatmap overlays visualizing the tile-based explainability attribution from ABMIL and our proposed ABMIL-IB architecture. Both trained models correctly classified the patient as belonging to the longer-survival group. **a:** Raw WSI. **b:** Attention scores from ABMIL. Brighter regions highlight tiles that are considered to a greater extent in subsequent parts of the model. The WSI-level prediction is inferred by a subsequent classifier. **c:** Decision scores from our ABMIL-IB method. Red resp. blue regions highlight tiles associated with shorter resp. longer survival by the model. The color intensity highlights the degree of association with either survival group. White regions highlight tiles that are not associated with either survival group. The scores of all tiles are summed to obtain the WSI-level prediction.

## 2.4 Evaluation of prognostic value

Training and evaluation of the survival group classification model was performed using a stratified 5-fold cross-validation. Because the number of patients per hospital was relatively low, all patients in the shorter- and longer-survival groups from all three hospitals were combined. The cross-validation folds were drawn randomly while preserving the proportion of patients from each hospital and survival group. The discriminative performance of the model was assessed using standard metrics for binary classification, including area under the operator receiver curve (AUROC) and the F1 score. The predictions from the test folds of the cross-validation were combined, and the metric scores were calculated from all predictions. Hospital-specific metric scores were calculated based solely on respective patient predictions. For the AUROC scores, 95% confidence intervals were estimated using the Hanley-McNeil formula[43]. For comparison, state-of-the-art MIL architectures—including ABMIL, AdditiveMIL, and TransMIL[44]—were trained and evaluated using the same training configurations and cross-validation folds (all training details in Supplementary Table 2).

Next, we evaluated the prognostic value of the trained models on the entire study population, also including the intermediate survival group. For this purpose, we treated the groups predicted by the model as two risk groups. For patients in the shorter and longer survival groups, predictions were taken from the respective test folds of the cross-validation. For patients in the intermediate survival group who were not included in the cross-validation, predictions were generated by randomly applying one of the trained cross-validation models per patient. The difference in overall survival time between the risk groups was assessed using Kaplan-Meier curves and Cox proportional hazards regression. The regression analysis adjusted for patient age, sex and MGMT promoter methylation status. Patients with unknown MGMT status were excluded from this analysis.

## 2.5 Sampling of prognosis-relevant tiles

Prognosis-relevant tiles were randomly sampled based on their respective test fold decision scores. Larger negative or positive decision score values indicate a stronger association with shorter or longer survival, respectively. The decision scores depend solely on the FM embeddings of the given tiles and their interpretation is not influenced by the overall accuracy of the derived WSI-level survival group prediction. All tiles from all WSIs of a particular hospital were grouped into two sets, depending on whether their decision scores were positive or negative. From each set, 100,000 tiles were randomly sampled with replacement. Sampling was weighted according to the absolute values of the decision scores, so that tiles with stronger association with shorter or longer survival were more likely to be sampled. This process was performed for each hospital independently, resulting in a total of 600,000 sampled tiles.

## 2.6 SAE-based visual pattern mapping

Next, the prognosis-relevant tiles were mapped to human-interpretable and GBM-IDHwt-specific visual patterns. For this task, we trained a SAE using the Virchow2 embeddings of 21,193,639 tiles extracted from 1878 WSIs of H&E-stained FFPE glioblastoma tissue sections from five public data collections (CPTAC-GBM[45], DBTA[46], TCGA-GBM[28,29], TCGA-LGG[30], UPENN-GBM[47]). The WSIs were processed in the same manner as described in Subsection 2.2. The SAE was trained using public data collections to ensure that visual patterns were learned independently from the study dataset.

Like regular autoencoders, SAEs consist of an encoder and a decoder and they are trained to reconstruct the input embeddings as linear combinations of learned basis vectors. The coefficients and the basis vectors of the linear combinations are inferred by the encoder and decoder, respectively. In our study, the basis vectors correspond to visual patterns and the coefficients quantify their respective activations. Two important distinctions enable SAEs to learn visual patterns that are particularly easy to understand by humans: A large number of latent dimensions, N, which is typically much greater than the dimensionality of the input embeddings, and a sparsity constraint, that enforces the linear combinations to include only a sparse subset of visual patterns. In this study, we used a TopK-SAE[48], which implements a particularly simple and effective sparsity constraint: Each embedding is encoded as a linear combination of the K most activated visual patterns. As parameters we choose N = 7680 and K = 8 (all training details in Supplementary Table 3).

## 2.7 Selection of relevant visual patterns

After SAE training, the encoder was used to map the sampled prognosis-relevant tiles to sparse linear combinations of the learned visual patterns. We compared the coefficients of the obtained linear combinations between the sampled tiles associated with longer and shorter survival. The coefficients indicate how strongly the respective visual patterns were activated among the sampled tiles. To assess the difference in visual pattern activation between longer and shorter survival, the activations for each of the 7680 visual patterns were averaged for each survival group individually and the differences between the groups were calculated. To enable a comparison across hospitals, this was performed for each hospital independently, yielding one difference value for each visual pattern and hospital. The 24 visual patterns with the greatest mean absolute difference values across all hospitals were selected for further analysis. Beyond these 24 patterns the differences in activation between both survival groups were only marginal and not considered meaningful. Lastly, the vector representations of the selected 24 visual patterns in the FM embedding space were visualized using the PacMAP algorithm[49] with default parameters.

## 2.7 Morphological feature interpretation

The final interpretation of the relevant visual patterns was performed by three board-certified neuropathologists with longstanding experience in diagnostic evaluation of brain tumors. Interpretation involved mapping the visual patterns to meaningful histomorphological features. For every relevant pattern, the neuropathologists were presented with randomly sampled example tiles (available at https://doi.org/10.5281/zenodo.18324224). Sampling was done without replacement and weighted according to the tiles' respective visual pattern activation. Thus, tiles with larger activations were more likely to be sampled. This process was performed for each hospital independently. For each of the three hospitals and the 24 visual patterns, 35 example tiles were sampled from the hospitals' 200,000 prognosis-relevant tiles, resulting in a total of 2520 example tiles. In an initial joint inspection, the neuropathologists found that the example tiles represented nine histomorphological categories: "tumor low cell-density", "tumor medium cell-density", "tumor high cell-density", "infiltration zone (low tumor-cell density)", "infiltration zone (high tumor-cell density)", "hemorrhage", "necrosis", "mixed", and "other". Thereafter, each neuropathologist independently evaluated the example tiles, assessing whether they exhibited meaningful histomorphological features that were consistent across hospitals and, if applicable, assigning them to the nine categories.

# 3. Results

## 3.1 Statistical characteristics of patients

The dataset compiled for this study included 720 GBM-IDHwt patients in total, with considerably more patients from Hospital 1 and Hospital 2 compared to Hospital 3 (Table 2). Across all hospitals, key statistical characteristics, including median age, distribution of sex, median survival time and, with the exception of Hospital 3, MGMT promoter methylation status were comparable to those reported by other population-based cancer registry studies[1,5,50–53]. The shorter- and longer-survival groups included similar patient numbers in each hospital.

| | **Hospital 1** | **Hospital 2** | **Hospital 3** | **Total** |
|---|---|---|---|---|
| **Number of patients** | 242 | 335 | 143 | 720 |
| ≤180 days | 90 | 130 | 51 | 271 |
| >180 and <360 days | 58 | 80 | 35 | 173 |
| ≥360 days | 93 | 112 | 57 | 262 |
| censored <360 days | 1 | 13 | 0 | 14 |
| **Sex [female / male]** | 116 / 126 | 127 / 208 | 59 / 84 | 302 / 418 |
| ≤180 days | 48 / 42 | 53 / 77 | 22 / 29 | 123 / 148 |
| >180 and <360 days | 25 / 33 | 26 / 54 | 14 / 21 | 65 / 108 |
| ≥360 days | 43 / 50 | 40 / 72 | 23 / 34 | 106 / 156 |
| **Median age* [year (IQR)]** | 67.5 (57.5–77.5) | 67.5 (57.5–77.5) | 67.5 (57.5–77.5) | 67.5 (57.5–77.5) |
| ≤180 days | 72.5 (67.5–77.5) | 77.5 (67.5–82.5) | 72.5 (67.5–77.5) | 72.5 (67.5–77.5) |
| >180 and <360 days | 67.5 (57.5–77.5) | 67.5 (62.5–77.5) | 67.5 (62.5–72.5) | 67.5 (57.5–77.5) |
| ≥360 days | 62.5 (52.5–67.5) | 62.5 (57.5–67.5) | 62.5 (52.5–67.5) | 62.5 (52.5–67.5) |
| **Median survival time [days]** | 274 | 251 | 302 | 269 |
| ≤180 days | 93 | 70 | 75 | 76 |
| >180 and <360 days | 269 | 270 | 269 | 269 |
| ≥360 days | 636 | 540 | 663 | 591 |
| **MGMT** [unmethylated / methylated]** | 144 / 93 | 167 / 124 | 50 / 93 | 361 / 310 |
| ≤180 days | 53 / 34 | 53 / 50 | 16 / 35 | 122 / 112 |
| >180 and <360 days | 40 / 17 | 49 / 20 | 13 / 22 | 102 / 59 |
| ≥360 days | 50 / 42 | 61 / 45 | 21 / 36 | 132 / 122 |

**Table 2:** Statistical characteristics of patients described in Subsection 2.1. The rows under each item stratify the patients into disjoint groups based on the number of days they survived after their initial diagnosis. The shorter- resp. longer-survival groups described in Subsection 2.3 are denoted by ≤180 days and ≥360 days, respectively. For conciseness, statistics for patients who were censored <360 days are omitted.

* = Median age was calculated using 5-year age groups; as a result, the median age was identical across all hospitals. ** = Patients with unknown MGMT methylation status exist.

## 3.2 Evaluation of prognostic value

Based on the results of the cross-validation, all of the compared MIL architectures, except for AdditiveMIL, demonstrated some capability to distinguish between patients with longer and shorter survival times (Table 3). This is evidenced by the AUROC scores and the corresponding 95% confidence intervals, which did not include the value 0.5. Corresponding F1 scores are given in the Supplementary Table 1. Our ABMIL-IB method performed similarly to the original ABMIL. TransMIL performed better, albeit not by a wide margin. A clear tendency toward comparable performance in Hospitals 1 and 2, and a near-random performance in Hospital 3 was observed across all MIL architectures, except for AdditiveMIL.

| **MIL architecture** | **Number of learnable parameters** | **Overall metric score** | **Hospital-specific metric scores** | | |
|---|---|---|---|---|---|
| | | | **Hospital 1** | **Hospital 2** | **Hospital 3** |
| ABMIL | 361 K | 0.66 (0.62–0.71) | 0.72 (0.65–0.80) | 0.72 (0.66–0.79) | 0.40 (0.29–0.51) |
| ABMIL-IB | 361 K | 0.67 (0.63–0.72) | 0.73 (0.65–0.80) | 0.72 (0.66–0.79) | 0.44 (0.33–0.55) |
| AdditiveMIL | 361 K | 0.49 (0.45–0.54) | 0.47 (0.38–0.56) | 0.51 (0.43–0.58) | 0.51 (0.40–0.62) |
| TransMIL | 2.8 M | 0.71 (0.67–0.76) | 0.74 (0.67–0.81) | 0.75 (0.69–0.81) | 0.53 (0.42–0.64) |

**Table 3:** Cross-validation AUROC scores of WSI-based survival group classification. The predictions from the test folds were combined and the AUROC scores were calculated from all predictions. The 95% confidence intervals are specified in brackets.

The results from the survival analysis of the predicted risk groups for the entire study population are displayed in Figure 3. The Kaplan-Meier curves of both groups were well-separated with median survival times of 216 and 350 days for the higher- and lower-risk group, respectively. Multivariate Cox proportional hazards regression adjusted for patient age, sex and MGMT promoter methylation status confirmed a difference in survival time (hazard ratio: 1.47; 95% confidence interval: 1.26–1.72).

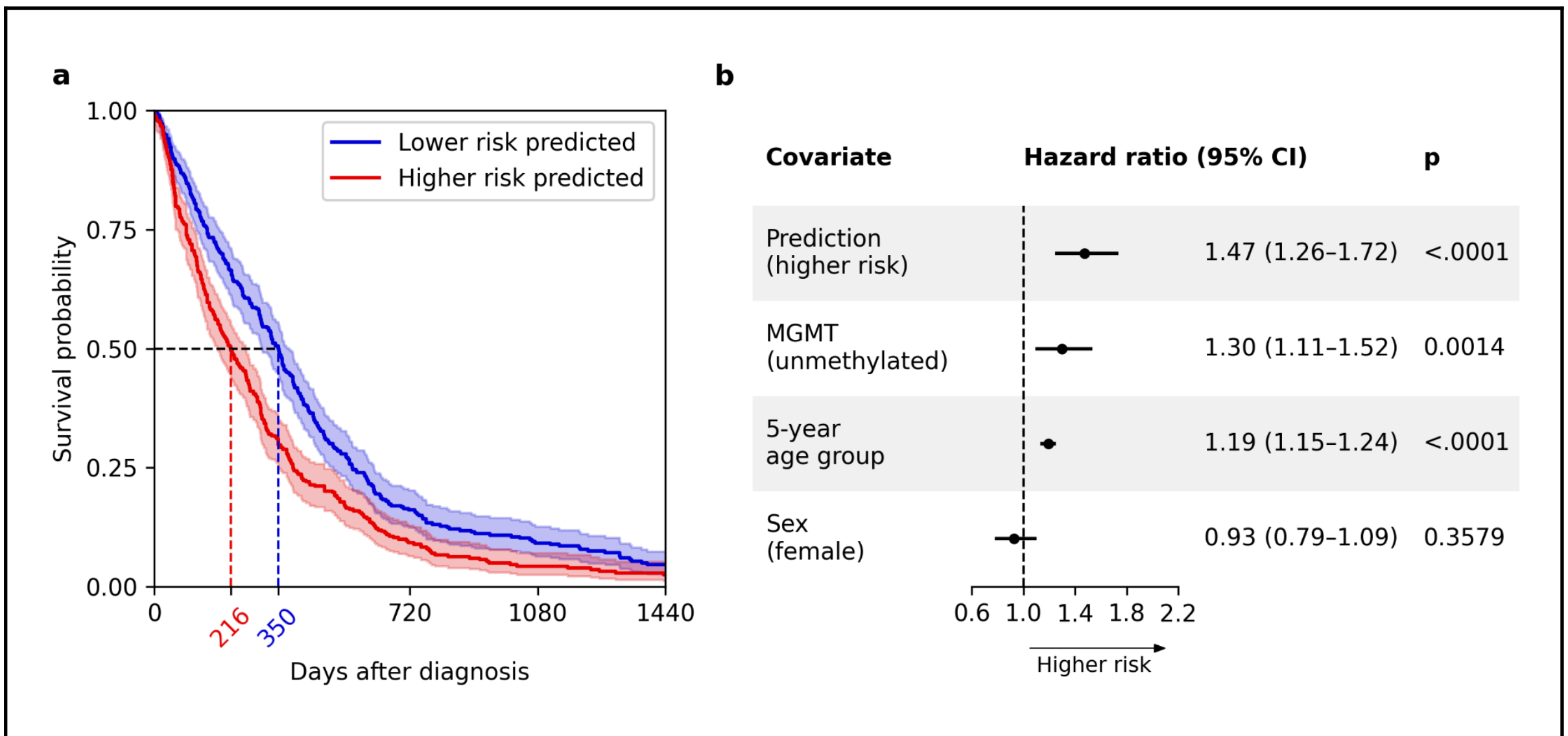

**Figure 3:** Survival analysis of the predicted risk groups. All patients were considered. **a:** Kaplan-Meier curves of the predicted lower- and higher-risk groups. Curves were cut off after 1440 days. **b:** Results from multivariate Cox proportional hazards regression.

## 3.3 SAE-based pattern mapping

From the 24 visual patterns with the largest mean absolute difference in activation across all hospitals, 11 were more activated among the prognosis-relevant tiles associated with longer survival and 13 were more activated among the prognosis-relevant tiles associated with shorter survival (Figure 4 Panel a). For each pattern the magnitude of the difference in activation and the direction of the survival association were mostly similar across all hospitals. The PacMAP visualization of the relevant visual patterns' spatial arrangement, indicates a clear separation between the patterns associated with shorter survival and those associated with longer survival in the FM embedding space (Figure 3 Panel c). This suggests that the patterns represent different visual concepts, which is evident when examining the example tiles sampled for each pattern (Figure 4 Panel b). The patterns associated with shorter survival formed two clearly defined clusters, suggesting two visually different concepts.

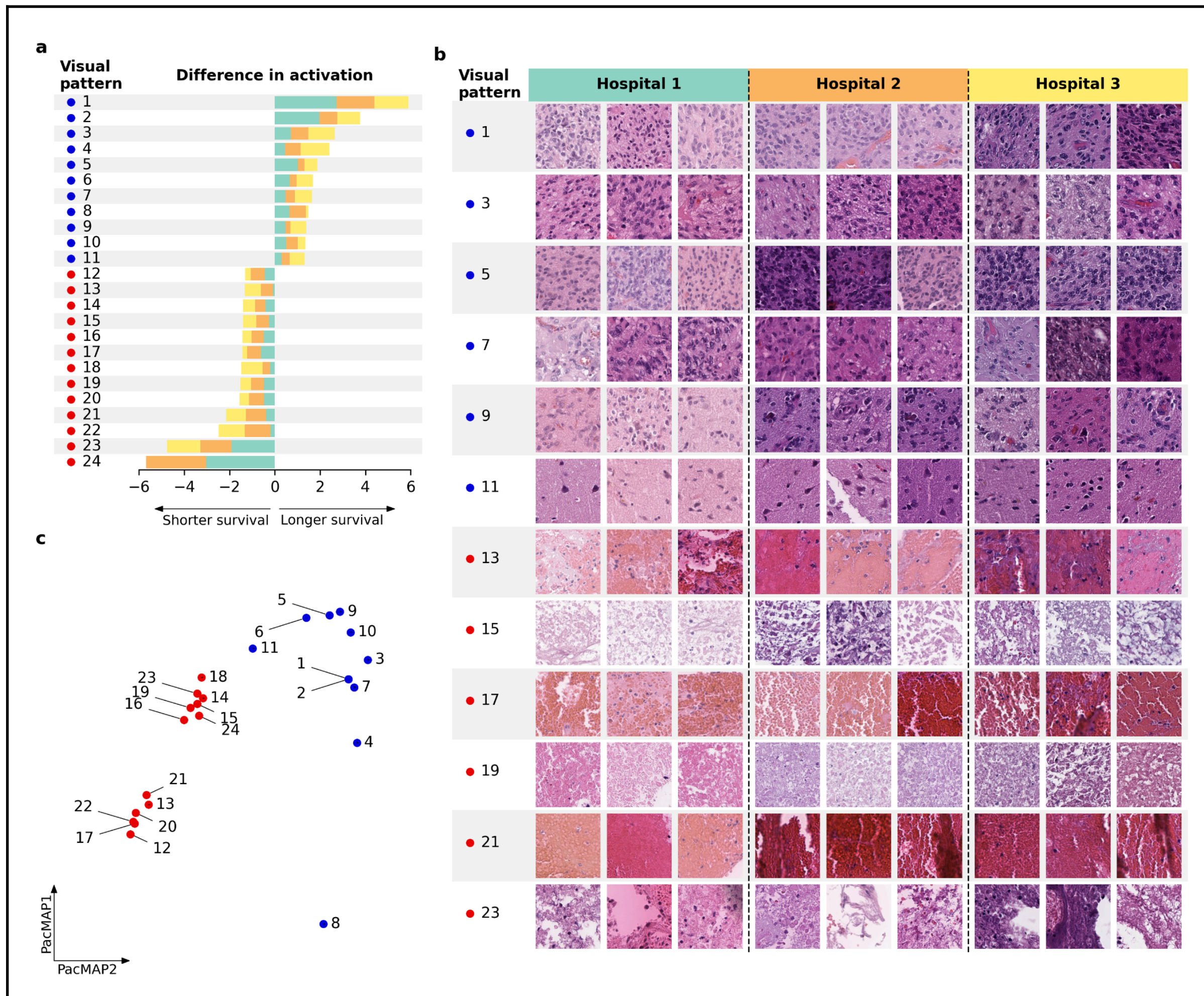

**Figure 4:** Visual patterns associated with shorter and longer survival. Blue and red markers indicate visual patterns that were more activated among tiles strongly associated with longer or shorter survival, respectively. **a:** Differences in the activation of visual patterns between the predicted longer and shorter survivors of Hospital 1 (turquoise), Hospital 2 (orange), and Hospital 3 (yellow). The 24 visual patterns with the greatest mean absolute difference in activation across hospitals are shown and ordered by their mean difference values. **b:** Example tiles for every second visual patterns from each hospital. **c:** The spatial arrangement of the 24 relevant visual patterns visualized using the PacMAP algorithm. Each point represents one visual pattern. Points closer to each other represent more similar visual patterns.

## 3.4 Morphological feature interpretation

The three neuropathologists found that 21 of the 24 selected visual patterns could each be clearly attributed to a uniform histomorphological category. The remaining three patterns were interpreted as mixtures of a few visually similar categories (Figure 5). The interpretations of the patterns were consistent across all hospitals despite notable differences in the tiles' visual appearance most likely originating from hospital-related differences in the preparation and digitization of the tissue. All patterns were attributed to seven histomorphological categories in total. The neuropathologists agreed on 13 visual patterns and disagreed only on the degree of tumor cell-density in five additional patterns. A disagreement between low cell-density tumor and infiltration zone was observed in patterns 9, 14, and 18. Overall, necrosis and hemorrhage was found to be associated with shorter survival, while cell-dense tumor and infiltration zone with low tumor cell-density were attributed to longer survival.

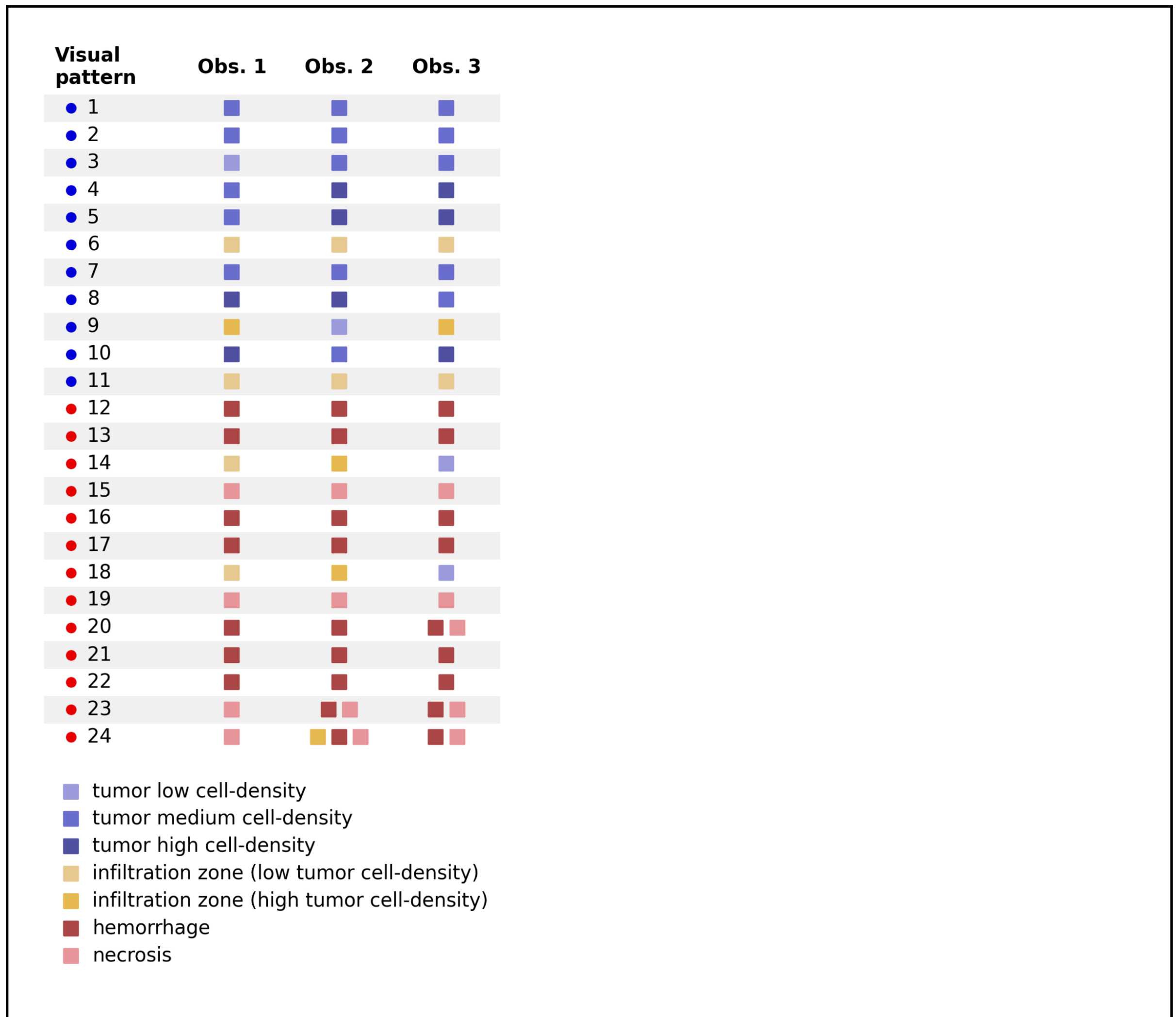

**Figure 5:** Mapping of relevant visual patterns to histomorphological features by three neuropathologists (Observer 1–3).

# 4. Discussion

## 4.1 Medical aspects

The evaluation showed that our method is capable of capturing interpretable associations between GBM-IDHwt histomorphology and patient survival. Similar to previous studies, we treated survival prediction as a classification task of two well-separated survival groups comprising 70% of the entire study population[17,25]. Subsequent survival analysis with the entire study population revealed that the WSI-based model predictions were an independent prognostic factor, separate from other established factors such as patient age and MGMT promoter methylation status. As expected, considering the long history of attempts to identify prognostic markers in GBM, the strength of the prognostic association was limited and similar in magnitude to that observed in previous studies[17,25–27].

Our work differs from many previous studies in that we invested considerable effort in creating a new real-world dataset consisting of linked pathology and population-based cancer registry data. This dataset contains 720 cases from three hospitals, which were selected based on consistent inclusion criteria. Therefore, it is much larger and more representative of the real-world patient population than previous datasets, most of which were originally compiled for different purposes, mostly clinical trials with patients selected by certain inclusion criteria. Thus, our study provides strong evidence for the prognostic value of GBM-IDHwt histomorphology.

The model's substantially lower ability to discriminate survival groups in Hospital 3 may be due to obvious differences in image appearance compared to Hospital 1 and Hospital 2 resulting from variations in tissue preparation and digitization. The considerably lower proportion of patients from Hospital 3 may also limit the model's capacity to adapt to domain-specific characteristics. Still, interpretations of visual patterns were consistent across all hospitals.

Our framework identified both established and unexpected associations between histomorphology and survival. Regarding shorter survival, necrosis and hemorrhage appeared to be the most relevant histomorphological features. These associations are well-known and were also demonstrated for GBM-IDHwt[25,26].

In terms of longer survival, highly cellular tumor areas and infiltration zones with low tumor cell density appeared to be most relevant. The former association is unexpected, as high cell density is known to be associated with malignant progression and poor prognosis in many tumor types. For example, high cellularity is one of the histopathological criteria defining high-grade astrocytoma, IDH-mutant (CNS WHO grade 3 and 4), as opposed to their low-grade counterpart (CNS WHO grade 2). However, histomorphological features such as hypercellularity are generally not considered to provide significant prognostic information in GBM-IDHwt[54]. Of note, tumor areas were mainly composed of fibrillary tumor cells, while gemistocytic or multinucleated tumor cells were usually spared and tumor cell nuclei lacked greater variances in size, shape or chromatin density. Recent findings based on manual search for histomorphological features also suggested tumor infiltration as well as other features indicative of tumor growth to be associated with longer survival, including cytologic atypia, mitoses, and hypercellularity[25,26]. One possible explanation is that cell-dense tumor areas represent tumor tissue that may be particularly susceptible to radiation or chemotherapy. Alternatively, the presence of these tissue areas on the slide could be attributed to confounding factors related to surgical practice, given that a greater extent of resection during surgery has previously been found to be associated with prolonged survival in glioblastoma[55,56].

## 4.2 Technical aspects

This is the first study to use SAEs for exploring the learned associations between histomorphology and patient survival in GBM-IDHwt. Our framework combines an explainable MIL architecture to identify prognosis-relevant tiles and a SAE to map these tiles post-hoc to interpretable visual patterns based on FM embeddings. This enables unbiased, quantitative analysis of the learned associations. It also makes it easier to identify potential confounders that would remain hidden in black-box models, or that could be obscured by manually searching for histomorphological features in attention heatmaps of selected cases. SAEs have significant advantages over cluster-based mapping approaches: They are less sensitive to parameter choices and can be efficiently trained to learn relevant visual patterns from very large histopathological datasets containing millions of tiles[22,23]. The TopK-SAE architecture used in this study has only two parameters specifying the upper bounds of the number of visual patterns that can be learned from the training data and the number of visual patterns used to encode each tile-based FM embedding. Tests with different values for the parameters N and K always resulted in the selection of the same types of histomorphological patterns.

The presented explainable AI framework is not limited to GBM-IDHwt and may facilitate prognostic biomarker discovery across a broader range of tumor entities. With respect to GBM-IDHwt specifically, we are releasing the trained SAE model to enable standardized quantitative comparison of visual patterns across future studies (available at https://doi.org/10.5281/zenodo.18324224). The model can be used out-of-the-box in future studies. As the SAE was trained in an unsupervised manner using image data alone, it is not restricted to survival prediction and also applicable to other GBM-IDHwt-related research questions.

Our proposed ABMIL-IB architecture for survival group prediction is an instance-based MIL variation of ABMIL. It achieved similar performance to the original embedding-based version while offering improved explainability. Both architectures use attention mechanisms to quantify the prognostic relevance of each tile based on its FM embedding. In ABMIL, the attention-weighted tile embeddings are first aggregated into an intermediate representation, which is then processed by a subsequent classifier to infer the final prediction. This makes it difficult to retrace the evidence individual tiles provide for the predicted survival group. In contrast, the ABMIL-IB architecture explicitly assigns a decision score to each tile indicating its association with either shorter or longer survival. The final WSI-level prediction is simply the sum of the normalized decision scores, making the evidence each tile provides fully transparent. The normalization of the decision scores is based on the tile-level attention scores across the entire WSI. This ensures that the WSI-level prediction remains unaffected by the number of tiles of specific histomorphological patterns. Compared to TransMIL, the proposed architecture performed slightly worse. However, TransMIL is a much more complex MIL architecture that is difficult to interpret. While this approach has the potential to model more complex spatial patterns across tiles, it only slightly improved performance compared to ABMIL-IB. This may be explained by the limited spatial structure of the sampled tissue, resulting from surgical excision and preparation.

It might seem reasonable to further improve explainability by performing MIL directly based on the visual pattern activations of individual tiles derived from SAE rather than their raw FM embeddings. However, it was previously reported that this does not improve performance[22], and our own empirical tests showed substantially weaker performance when training with visual pattern activations. SAEs are designed so that the dimensionality of the visual pattern activations is much larger than that of the original FM embeddings. Using these activations as input for the MIL model leads to a considerably larger number of learnable parameters, potentially making the model prone to overfitting.

## 4.3 Limitations and future work

The present work is subject to a number of limitations. First, the datasets' focus on Germany restricts the generalizability of the findings because population characteristics or treatment practices may vary in other countries. Additionally, as in most survival time analyses using real-world data, the potential for bias due to loss to follow-up cannot be entirely excluded. However, since the percentage of loss to follow-up was below 2%, we assume the influence to be negligible.

Second, as with any post hoc explainability approach, the identified histomorphological survival associations are only simplified approximations of a complex AI model's inner workings. Faithful interpretation of these associations requires a clear understanding of the applied methods and their limitations and may be affected by unintentional confirmation bias[14,15].

Third, the adoption of SAEs for mapping tiles to visual patterns is inherently limited. In the context of image analysis, no quantitative metrics have yet been established to evaluate or optimize the utility of the learned patterns. Consequently, the selection of SAE parameters in this work remains largely experimental. Furthermore, by design, SAEs produce a large number of visual patterns, making it time-consuming to assess each one systematically. To maintain feasibility, only the 24 most striking visual patterns were examined in detail. Finally, quantitative histological features, such as mitotic count, can only be inferred indirectly based on the tiles associated with each visual pattern.

For clinical translation, a better understanding of the association between GBM-IDHwt histomorphology and survival is required. In particular, the unexpected associations require closer examination, for instance by stratifying the evaluation according to clinical parameters that could be confounding factors, such as tumor location, extent of resection, and time of surgery. Equally important, future studies should investigate how GBM-IDHwt histomorphology can be integrated with established prognostic factors to improve prognostication accuracy. This will require larger and more diverse real-world datasets, covering multiple international hospitals.

From a technical point of view, future studies should explore how SAE training and parametrization affect the inferred visual patterns. It should also be investigated how the interpretation of visual patterns could be further automated and made more objective and quantitative, for instance, through visual-language foundation models for computational pathology[57].

# Acknowledgements

We would like to thank the four cancer registries who provided data for this study: Bremer Krebsregister, Klinisches Krebsregister Niedersachsen, Epidemiologisches Krebsregister Niedersachsen and Landeskrebsregister Nordrhein-Westfalen. This work was supported by the Center for Histology and Evaluation in Digital (Nephro) Pathology (HELP) at RWTH Aachen University.

# Author Contributions

JPR, FF, SN, STH, JW, SL, AE, and AH conceived the study. JPR, FF, CB, and AH designed the methodology. JPR, with support from AH, developed the software. Validation, formal analysis, and investigations were carried out by JPR, FF, STH, JW, CB, and AH. Data were provided and curated by JPR, FF, SN, NSS, STH, JW, SL, AE, and PB. FF, SN, STH, JW, SL, and AH provided supervision and project administration, and FF, SN, STH, JW, SL, AE, TI, KK, and AH secured funding. JPR and AH wrote the original draft of the manuscript. All authors contributed to writing, reviewing, and editing and approved the final paper.

# Data Availability

The use of the pathology and cancer registry data examined was only approved for this study. Therefore, these data cannot be made publicly available. However, interested researchers may contact the corresponding author for assistance in applying to use the data themselves. Study artifacts, including the example tiles presented to the neuropathologists and the model weights of the trained SAE, are released at https://doi.org/10.5281/zenodo.18324224. The code implementing the methods is released open-source at https://github.com/FraunhoferMEVIS/canconnect-study.

# Funding

The research reported in this publication was conducted in the context of the CanConnect project, supported by the German Federal Ministry of Health, based on a resolution of the German Bundestag (funding codes: ZMI5-2522DAT15A, ZMI5-2522DAT15B, ZMI5-2522DAT15C, ZMI5-2522DAT15D, ZMI5-2522DAT15E). FF received additional funding from ERACoSysMed and the German Federal Ministry of Education and Research (BMBF) under grant number FKZ 31L0237A (MiEDGE). PB is supported by the German Research Foundation (DFG, Project IDs 322900939 & 445703531 & INST 222/1582-1), European Research Council (ERC Consolidator Grant No 101001791), the Federal Ministry of Education and Research (BMBF, STOP-FSGS-01GM2202C), and the Innovation Fund of the Federal Joint Committee (Transplant.KI, No. 01VSF21048).

# Declaration of Competing Interest

The authors declare no competing interests.

# Ethics Approval and Consent to Participate

This study was approved by all ethics committees responsible for the three participating hospitals: Ethikkommission der Ärztekammer Bremen (No. 869), Ethikkommission der Medizinische Hochschule Hannover (No. 10888_BO_K_2023), and Ethikkommission des Universitätsklinikums Aachen (No. EK 23-133). The study was also approved by all four participating cancer registries: Bremer Krebsregister (Dec 20, 2023), Klinisches Krebsregister Niedersachsen (Sep 29, 2023), Epidemiologisches Krebsregister Niedersachsen (Dec 6, 2023) and Landeskrebsregister Nordrhein-Westfalen (Jan 2, 2024). Consent to participate was waived because the study relied exclusively on retrospective, anonymized data. Data processing complied with the national laws and regional regulations of the three German federal states from which the data was obtained.

# Supplementary material

## Privacy-preserving data linkage

The data provided by the three hospitals and the respective cancer registries was compiled, transferred and combined using privacy-preserving data linkage (Supplementary Figure 1). Initially, the cancer registries identified all patients who matched the inclusion criteria and pseudonymized each one with a new, unique study identifier (ID-S). Then, the registries provided the respective hospitals with mappings from the ID-S to the hospitals' original histology submission keys (ID-HSK). The registries also sent the requested registry data of the selected patients to the researchers, keyed only by the ID-S. Using the ID-HSK, the neuropathology departments of the hospitals compiled the pathology data for the selected patients. They then remapped this data to the ID-S and transferred it to the researchers. The researchers linked the pathology and registry data of individual patients using only the ID-S. This data linkage process ensured that the researchers did not receive any internal patient identifiers from the hospitals or cancer registries.

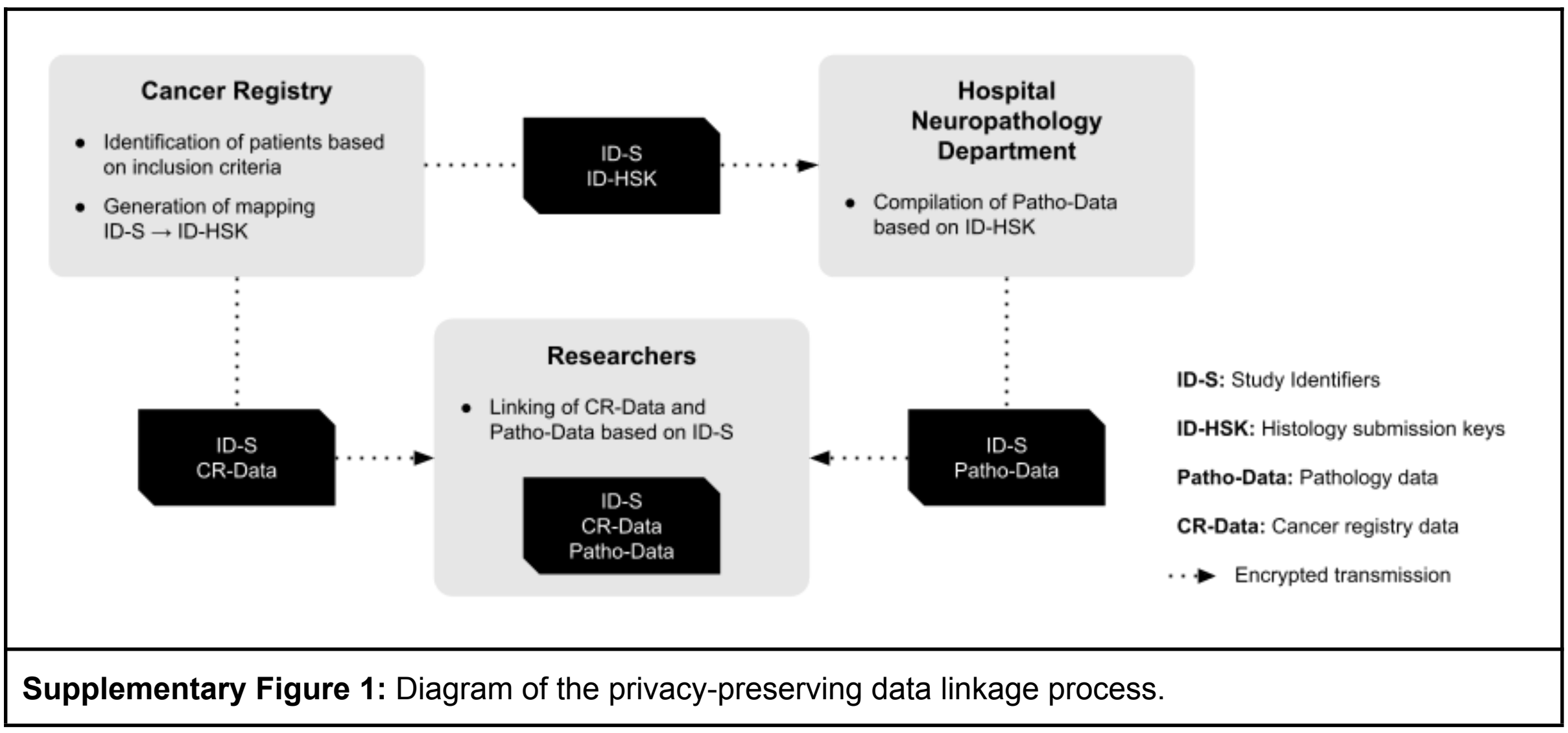

**Supplementary Figure 1:** Diagram of the privacy-preserving data linkage process.

## Evaluation of prognostic value: F1 scores

| MIL method | Number of model parameters | Overall score | Hospital-specific scores | | |
|---|---|---|---|---|---|
| | | | Hospital 1 | Hospital 2 | Hospital 3 |
| ABMIL | 361 K | 0.63 | 0.69 | 0.66 | 0.46 |
| ABMIL-IB | 361 K | 0.62 | 0.68 | 0.63 | 0.50 |
| AdditiveMIL | 361 K | 0.54 | 0.55 | 0.53 | 0.57 |
| TransMIL | 2.8 M | 0.64 | 0.68 | 0.66 | 0.53 |

**Supplementary Table 1:** Cross-validation F1 scores of WSI-based survival group classification. The predictions from the test folds were combined and the F1 scores were calculated from all predictions.

# Details on the training of AI-based methods

| Item | Specification |
| --- | --- |
| Loss function | Binary cross entropy loss |
| Optimizer | AdamW |
| Initial learning rate | 5 x 10e-5 |
| Learning rate scheduler | Cosine annealing without warmup |
| Weight decay | 1 x 10e-2 |
| Batch size | 1 |
| Gradient accumulation steps | 32 |
| Epochs | 16 |
| GPU | NVIDIA RTX A5000 24 GB |

**Supplementary Table 2:** Details of the MIL-based survival group classification training.

| Item | Specification |
| --- | --- |
| Number of latent dimensions | 5120, 7680, 10240 |
| Sparsity constraint K | 4, 8, 12 |
| Loss function | Reconstruction loss (no auxiliary loss) |
| Optimizer | Adam |
| Initial learning rate | 5 x 10e-5 |
| Learning rate scheduler | Linear decay during the last 20% of training steps |
| Batch size | 16384 |
| Epochs | 8 |
| GPU | NVIDIA RTX A5000 24 GB |

**Supplementary Table 3:** Details of the SAE training.

| Data collection | Number of WSIs | Number of tiles | Case selection criteria | Link to data collection |
| --- | --- | --- | --- | --- |
| CPTAC-GBM[45] | 224 | 852,158 | IDH-wildtype | https://www.cancerimagingarchive.net/collection/cptac-gbm/ |
| DBTA[46] | 662 | 7,997,959 | Glioblastoma IDH-wildtype, astrocytoma IDH-wildtype, | https://search.kg.ebrains.eu/instances/Dataset/8fc108ab-e |

| | | | giant cell glioblastoma, gliosarcoma, control | 2b4-406f-8999-60269dc1f994 |
|---|---|---|---|---|
| TCGA-GBM[28,29] | 790 | 9,721,966 | IDH-wildtype | https://portal.gdc.cancer.gov/projects/TCGA-GBM |
| TCGA-LGG[30] | 131 | 1,629,487 | IDH-wildtype | https://portal.gdc.cancer.gov/projects/TCGA-LGG |
| UPENN-GBM[47] | 71 | 992,069 | all cases | https://www.cancerimagingarchive.net/collection/upenn-gbm/ |
| **Total** | **1878** | **21,193,639** | | |

**Supplementary Table 4:** Overview of the datasets used for the SAE training.